\documentclass[12pt,preprint]{article}
\usepackage{amsmath}
\usepackage{graphics}
\usepackage{epsfig}
\usepackage{psfig}
\renewcommand{\vec}[1]{{\bf #1}}
\normalsize
\newcommand{\eqb}{\begin{equation}}
\newcommand{\eqe}{\end{equation}}
\newcommand{\dmb}{\begin{displaymath}}
\newcommand{\dme}{\end{displaymath}}
\newcommand{\pd}{\partial}

\newcommand{\eab}{\begin{eqnarray}}
\newcommand{\eae}{\end{eqnarray}}

\newcommand{\be}{\begin{equation}}
\newcommand{\ee}{\end{equation}}

\begin{document}
\begin{titlepage}
\begin{flushright}
2004-11\\ 
HD-THEP-04-46
\end{flushright}
\vspace{0.6cm}

\begin{center}
\Large{Emergent inert adjoint scalar field 
in SU(2) Yang-Mills thermodynamics due to 
coarse-grained topological fluctuations}

\vspace{1cm}

\large{Ulrich Herbst and Ralf Hofmann$\mbox{}^\dagger$}

\end{center}
\vspace{0.3cm}

\begin{center}
{\em 
Institut f\"ur Theoretische Physik\\ 
Universit\"at Heidelberg\\ 
Philosophenweg 16\\ 
69120 Heidelberg, Germany}
\end{center}
\vspace{0.3cm}
\begin{center}
{\em $\mbox{}^\dagger$Institut f\"ur Theoretische Physik\\ 
Universit\"at Frankfurt\\ 
Johann Wolfgang Goethe - Universit\"at\\ 
Max von Laue -- Str. 1\\ 
D-60438 Frankfurt, Germany}
\end{center}
\vspace{0.5cm}
\newpage 

\begin{abstract}

We compute the  
phase and the modulus of an energy- and pressure-free, composite, adjoint, and inert 
field $\phi$ in an SU(2) Yang-Mills theory at large temperatures. 
This field is physically relevant in describing part of the 
ground-state structure and the quasiparticle masses of excitations.
The field $\phi$ possesses 
nontrivial $S^1$-winding on the group 
manifold $S^3$. Even at asymptotically high temperatures, where the 
theory reaches its Stefan-Boltzmann limit, the field $\phi$, 
though strongly power-suppressed, is 
conceptually relevant: its presence resolves the infrared 
problem of thermal perturbation theory.

\end{abstract} 

\end{titlepage}

\section{Introduction}

In \cite{Hofmann2004} one of us has put 
forward an analytical and nonperturbative approach 
to SU(2)/SU(3) Yang-Mills thermodynamics. 
Each of theses theories comes in three phases: 
deconfining (electric phase), preconfining (magnetic phase), and completely confining (center phase).
This approach 
assumes the existence of a composite, adjoint Higgs field 
$\phi$, describing part of the thermal 
ground state, that is, the BPS saturated topologically nontrivial sector 
of the theory. The field $\phi$ is generated 
by a spatial average over noninteracting trivial-holonomy SU(2) calorons \cite{HarrigtonShepard1977} which can be 
embedded in SU(3). 
The `condensation'\footnote{By `condensation' we mean the effects of long-range spatial correlations
in the classical, BPS saturated, trivial-holonomy configurations 
in singular gauge, see (\ref{defphi}).} of trivial-holonomy SU(2) calorons into the field $\phi$ 
must take place at an asymptotically high temperature \cite{Hofmann2004}, that is, at the 
limit of applicability of the gauge-field theoretical description. 
For any physics model formulated in terms of 
an SU(2)/SU(3) Yang-Mills theory this is to say that caloron 'condensation' takes place 
at $T\sim M_P$ where $M_P$ denotes the Planck mass. Since 
$|\phi|\sim\sqrt{\frac{\Lambda^3}{2\pi T}}$ topological defects only very marginally 
deform the ideal-gas expressions for thermodynamical quantities at 
$T\gg\Lambda$. Here 
$\Lambda$ denotes the Yang-Mills scale. Every contribution to a thermodynamical quantity, which 
arises from the topologically nontrivial 
sector, is power suppressed in temperature. As a consequence, 
the effective theory is {\sl asymptotically free} and exhibits the same infrared-ultraviolet 
decoupling property \cite{Hofmann2004} that is seen in renormalized perturbation theory \cite{'thooft}. 
Asymptotic freedom is a conceptually very 
appealing property of SU(N) Yang-Mills
theories. It first was discovered in 
perturbation theory \cite{NP2004}. 

In the effective thermal theory interactions between 
trivial-holonomy calorons in the ground state are taken 
into account by obtaining a pure-gauge solution to 
the classical equations of motion for the 
topologically trivial sector in the (nonfluctuating and nonbackreacting) background 
$\phi$. Thus the partition function of the fundamental theory 
is evaluated in three steps in the electric phase: (i) integrate over the admissible part of the moduli
space for the caloron-anticaloron system and spatially average over the associated two-point correlations 
to derive the (classical and 
temperature dependent) dynamics of an adjoint, spatially homogeneous scalar field 
$\phi$, (ii) establish the quantum mechanical and statistical 
inertness of $\phi$ and use it as a temperature dependent background to find a 
pure-gauge solution $a^{bg}_\mu$ to the Yang-Mills equations describing 
the trivial-topology sector. Together, $\phi$ and $a^{bg}_\mu$ 
constitute the thermal ground state of the system. The fact that the 
action for the ground-state configurations $\phi$ and $a^{bg}_\mu$ is 
infinite is unproblematic since the corresponding, vanishing 
weight in the partition function 
is associated with a nonfluctuating 
configuration and therefore can be factored out and cancelled when evaluating expectation 
values in the effective theory. 
(iii) Consider the interactions between the macroscopic 
ground state and trivial-topology fluctuations in 
terms of quasiparticle masses of the latter which are generated by the adjoint 
Higgs mechanism\footnote{In unitary gauge off-Cartan fluctuations acquire a 
temperature dependent mass when propagating through and thereby interacting with the 'medium' 
of caloron fluctuations.} and 
impose thermodynamical selfconsistency to derive an 
evolution equation for the effective gauge coupling $e$. 
%***********************
\begin{figure}
\begin{center}
%\leavevmode
%%\epsfxsize=9.cm
%\leavevmode
%%\epsffile[80 25 534 344]{}
\vspace{5.3cm}
\includegraphics{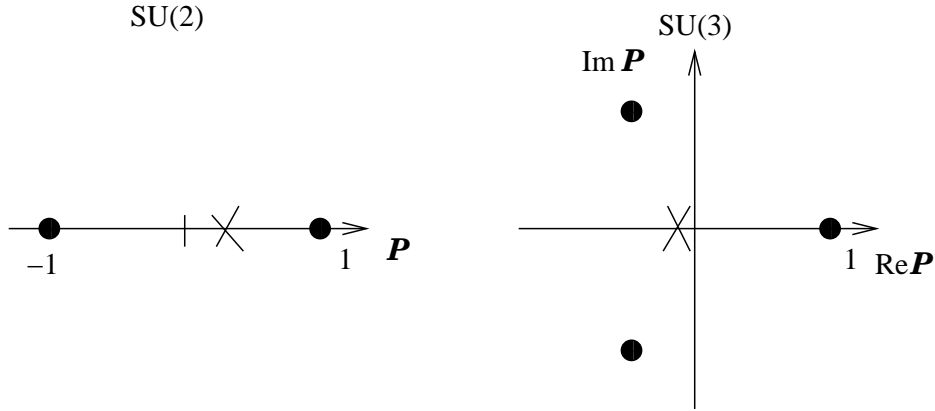}
\end{center}
\caption{Possible values of the Polyakov 
loop ${\cal P}$ at spatial infinity on a given gauge-field configuration. 
A small holonomy corresponds to values close to the center elements depicted by the 
dots. Crosses indicate examples 
for a large holonomy.\label{Polya}}      
\end{figure}
%************************ 
In the following we will restrict our discussion to the case SU(2). 
Isolated magnetic charges are generated by dissociating calorons of sufficiently large holonomy 
\cite{Nahm1984,LeeLu1998,KraanVanBaalNPB1998,vanBaalKraalPLB1998,Brower1998,
Diakonov2004,KorthalsAltes,HoelbingRebbiRubakov2001}, for a quick explanation of the term 
holonomy see Fig.\,\ref{Polya}. Nontrivial holonomy is locally excited 
by interactions between trivial-holonomy calorons and 
anticalorons. In \cite{Diakonov2004} it was shown as a result of a heroic 
calculation that small (large) holonomy leads to an attractive 
(repulsive) potential between the monopole and the antimonopole constituents 
of a given caloron. An attraction between a monopole and an 
antimonopole leads to annihilation if the distance between their 
centers is comparable to the sum of 
their charge radii. Thermodynamically,
the probability for exciting a small holonomy is much larger than that for exciting a
large holonomy. In the former case this probability roughly is determined
by the one-loop quantum weight of a trivial holonomy caloron,
while in the latter case both monopole constituents have a combined mass
$\sim 4\pi^2 T\sim 39\,T$ \cite{LeeLu1998}. 
Thus an attractive potential 
between a monopole and its antimonopole is the 
by far dominating situation. This is the microscopic reason 
for a negative ground-state pressure $P^{g.s.}$ which, on spatial average, turns out to 
be $P^{g.s.}=-4\pi\Lambda^3T$ (the equation of ground state is 
$\rho^{g.s.}=-P^{g.s.}$) \cite{Hofmann2004}. In the unlikely case of repulsion 
(large holonomy) the monopole and the antimonopole separate 
back-to-back until their interaction is sufficiently 
screened to facilitate their existence in 
isolation (as long as the overall pressure of 
the system is positive). Magnetic monopoles in 
isolation do not contribute to the 
pressure of the system\footnote{The reader may convince himself of this fact 
by computing the energy-momentum tensor on a 
BPS monopole.}. The overall pressure is positive if the gauge-field 
fluctuations after spatial coarswe-graining are sufficiently light and 
energetic to over-compensate the negative ground-state contribution, 
that is, if the temperature is sufficiently large.      

Caloron-induced tree-level 
masses for gauge-field modes decay as  
$1/\sqrt{T}$ when heating up the system. Due to the linear rise of $\rho^{g.s.}$ with $T$ 
the thermodynamics of the 
ground state is thus subdominant at large temperatures\footnote{Gauge-field excitations 
are free at large temperatures \cite{HerbstHofmannRohrer2004} and 
contribute to the total pressure and the total energy density like $\sim T^4$. 
The small residual interactions, which peak close to a 2$^{\tiny\mbox{nd}}$-order
transition to the magnetic phase at $T_{c,E}$, 
are likely to explain the large-angle anomalies seen in some CMB power spectra 
\cite{HerbstHofmannRohrer2004,WMAP2003}. When cooling the system, 
monopoles and antimonopoles, which were generated 
by dissociating calorons (large holonomy), start to overlap at a temperature $T_o$ slightly higher 
than $T_{c,E}$ because they are moving towards one another under the 
influence of an overall negative pressure. The latter is 
generated by the dominating pressure component generated by 
small-holonomy calorons whose monopole constituents attract and 
eventually annihilate at a given point in 
space but get re-created elseswhere. Naively seen, 
negative pressure corresponds to an instability of the system causing it to 
collapse. We usually imagine a contracting system in terms of a 
decrease of the mean interparticle distance while tacitly assuming the particles 
to be pointlike. Despite an overall negative pressure a total collapse 
in the above sense does not occur in an SU(2) 
Yang-Mills theory. This can be understood as follows: 
The mass of an isolated, screened 
monopole is $m \sim \frac{2\pi^2 T}{e}$, and the effective gauge coupling $e$ 
is constant if monopoles do not overlap, that is, for $T>T_o$ 
(magnetic charge conservation \cite{Hofmann2004}). At $T_o$ the magnetic charge 
contained in a given spatial volume
no longer remains constant in time because of the increasing mobility of strongly screened monopoles.
Thus formerly separated monopoles can annihilate and but also get re-created.
Therefore the notion of a 
local collapse ceases to be applicable since the associated particles 
cease to exist if they are close to one another. If the rate of 
annihilation equals the rate of re-creation of monopole-antimonopole 
pairs then we are witnessing an equilibium situation characterized 
by a temperature despite a negative overall pressure.}.     
The main purpose of the present work is to compute and to discuss 
the dynamical generation 
of an adjoint, macroscopic, and composite scalar field $\phi$. This is a first-principle analysis 
of the ground-state structure in the 
electric phase of an SU(2) Yang-Mills theory. 
The paper is organized as follows: 
In Sec.\,\ref{phasephi} we write down and discuss a 
nonlocal definition, relevant for the determination of $\phi$'s phase, 
in terms of a spatial and scale-parameter average 
over an adjointly transforming $2$-point 
function. This average needs to be evaluated on trivial-holonomy 
caloron and anticaloron configurations at a 
given time $\tau$. In Sec.\,\ref{comphase} we 
perform the average and discuss 
the occurrence of a global gauge 
freedom in $\phi$'s phase, which has a geometrical interpretation. 
In Sec.\,\ref{macro} we show how 
the derived information about a nontrivial 
$S^1$ winding of the field $\phi$ together with analyticity of the 
right-hand side of the associated BPS equation and with the assumption of the existence of an externally given
scale $\Lambda$ can be used to 
uniquely construct a potential determining $\phi$'s classical (and temperature dependent) 
dynamics. 
In Sec.\,\ref{SO} we summarize and discuss our results 
and give an outlook on future research.

\section{Definition of $\phi$'s phase\label{phasephi}}

In this section we 
discuss the BPS saturated, topological part of the ground-state physics in the electric phase 
of an SU(2) Yang-Mills theory. 
According to the approach in \cite{Hofmann2004} the adjoint scalar 
$\phi$ emerges as an energy- and pressure-free (BPS saturated) field 
from a spatial average over the classical correlations in a caloron-anticaloron system 
of trivial holonomy in absence of interactions. On spatial average, the latter are taken 
into account by a pure-gauge configuration solving the classical, trivial-topology 
gauge-field equations in the spatially homogeneous background $\phi$. This is consistent 
since $\phi$'s quantum mechanical and statistical inertness can 
be established. Without assuming the existence of a Yang-Mills scale $\Lambda$
only $\phi$'s phase, that is $\frac{\phi}{|\phi|}$, 
can be computed. A computation of 
$\phi$ itself requires the existence of $\Lambda$.
As we shall see, the information about the $S^1$ 
winding of $\phi$'s phase together with the analyticity of the right-hand side of $\phi$'s BPS equation
uniquely determines $\phi$'s modulus in terms of 
$\Lambda$ and $T$. 

Let us first set up 
some prerequisites. We consider BPS saturated solutions to the Yang-Mills equation
%*********
\eqb
\label{YME}
D_\mu F_{\mu\nu}=0
\eqe
%********
which are periodic in Euclidean time 
$\tau$ ($0\le\tau\le\frac{1}{T}$) 
and of trivial holonomy.
The only relevant configurations are calorons of
topological charge\footnote{Configurations with higher topological 
charge and trivial holonomy have been constructed, see for example 
\cite{Actor1983,Chakrabarti1987}. A priori they should 
contribute to the ground-state thermodynamics of the theory in 
terms of additional adjoint scalar fields. 
The nonexistence of these Higgs fields 
is assured by their larger number of dimensionful moduli 
- for the charge-two case we have two instanton radii and the core separation - 
which does not allow for the nonlocal definition of a macroscopic, {\sl dimensionless} 
phase, see (\ref{defphi}) and the discussion following it.} $\pm 1$. 
They are \cite{HarrigtonShepard1977} 
%*********
\eab
\label{HS}
A^C_\mu(\tau,\vec{x})&=&\bar{\eta}_{a\mu\nu}\frac{\lambda^a}{2}\pd_{\nu}\ln \Pi(\tau,\vec{x})\,\ \ \ \mbox{or}\nonumber\\ 
A^A_\mu(\tau,\vec{x})&=&\eta_{a\mu\nu}\frac{\lambda^a}{2}
\pd_{\nu}\ln \Pi(\tau,\vec{x})\,
\eae
%********
where the 't Hooft symbols $\eta_{a\mu\nu}$ and $\bar{\eta}_{a\mu\nu}$ 
are defined as
%********
\eab
\label{tHooftsym}
\eta_{a\mu\nu} &=& \epsilon_{a\mu\nu} + \delta_{a\mu}\delta_{\nu4} - \delta_{a\nu}\delta_{\mu4} \nonumber\\ 
\bar\eta_{a\mu\nu} &=& \epsilon_{a\mu\nu} - \delta_{a\mu}\delta_{\nu4} + \delta_{a\nu}\delta_{\mu4}\,.
\eae
%**********
The solutions in Eq.\,(\ref{HS}) (the superscript (A)C refers to (anti)caloron) are 
generated by a temporal mirror sum of 
the 'pre'potential $\Pi$ of a single (anti)instanton in singular 
gauge \cite{Atiyah1978}. They have the same color orientation as 
the `seed' instanton or `seed' antiinstanton. 
In Eq.\,(\ref{HS}) $\lambda^a$, ($a=1,2,3$), denote the Pauli matrices. 
The 'nonperturbative'
definition of the gauge field is used were the gauge 
coupling {\sl constant} $g$ is absorbed into the field. 

\noindent The scalar function $\Pi(\tau,\vec{x})$ is given as \cite{HarrigtonShepard1977}
%***********
\eqb
\label{Pi}
\Pi(\tau,\vec{x})=\Pi(\tau,r)\equiv1+\frac{\pi\rho^2}{\beta r}
\frac{\sinh\left(\frac{2\pi r}{\beta}\right)}{\cosh\left(\frac{2\pi r}{\beta}\right)-
\cos\left(\frac{2\pi\tau}{\beta}\right)}\,,
\eqe
%*********
where $r\equiv|\vec{x}|$, $\beta\equiv 1/T $, and $\rho$ 
denotes the  scale parameter whose variation leaves the classical action $S=\frac{8\pi^2}{g^2}$ invariant. 
At a given $\rho$ the solutions in Eq.\,(\ref{HS})
can be generalized by shifting the center 
from $z=0$ to $z=(\tau_z,\vec{z})$ by the (quasi) 
translational invariance of the classical action\footnote{Because of periodicity, $\tau_z$ needs 
to be restricted as  $0\le\tau_z\le\beta$.} $S$.  Another set of moduli is 
associated with global color rotations of the solutions in Eq.\,(\ref{HS}). 
 
\begin{figure}
\begin{center}
%\leavevmode
%%\epsfxsize=9.cm
%\leavevmode
%%\epsffile[80 25 534 344]{}
\vspace{5.3cm}
\includegraphics{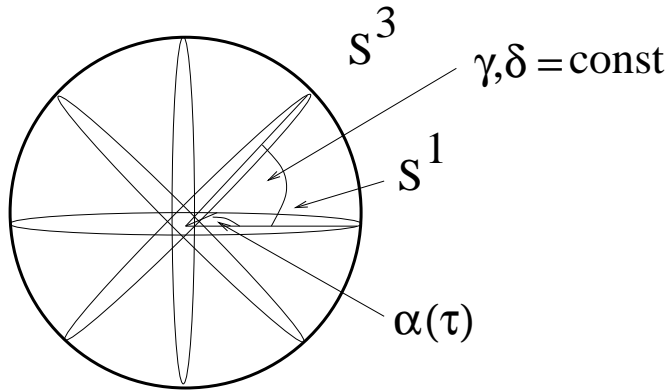}
\end{center}
\caption{Possible directions of winding of $\frac{\phi}{|\phi|}(\tau)$ 
around the group manifold $S^3$ of SU(2). The angles $\gamma,\delta$ 
are arbitrary but constant. They are determined by the choice of plane in which 
angular regularization is carried out when computing $\phi$'s phase, see below. 
The angle $\alpha(\tau)$ parametrizes 
the $S^1$ winding of $\frac{\phi}{|\phi|}$.\label{winding}}      
\end{figure}

From the BPS saturation
%*********
\eqb
\label{BPSCal}
F_{\mu\nu}[A^{(C,A)}]=(+,-)\tilde{F}_{\mu\nu}[A^{(C,A)}]
\eqe
%********
it follows that the (Euclidean) energy-momentum tensor 
$\theta_{\mu\nu}$, evaluated on $A_\mu^{(C,A)}$, vanishes identically
%********
\eqb
\label{thetacal}
\theta_{\mu\nu}[A^{(C,A)}]\equiv 0\,.
\eqe
%*********
This property translates to the macroscopic field 
$\phi$ with energy-momentum tensor $\bar{\theta}_{\mu\nu}$ in an effective 
theory since $\phi$ is obtained by a spatial average over caloron-anticaloron correlations 
neglecting their interactions\footnote{A spatial average over zero energy-momentum yields zero.}
%********
\eqb
\label{thetamacro}
\bar{\theta}_{\mu\nu}[\phi]\equiv 0\,.
\eqe
%********* 
The field $\phi$ is spatially homogeneous since it emerges from a spatial average. 
If the action density governing $\phi$'s dynamics in the absence of caloron 
interactions contains a kinetic term quadratic in the $\tau$-derivatives and a potential 
$V$ then Eq.\,(\ref{thetamacro}) is equivalent to $\phi$ solving the first-order equation
%********
\eqb
\label{BPSphi}
\pd_\tau\phi=V^{(1/2)}\,
\eqe
%********* 
where $V^{(1/2)}$ denotes the 'square-root' of $V$\footnote{The fact that an ordinary and not a 
covariant derivative appears in Eq.\,(\ref{BPSphi}) is, of course, tied to our specific gauge 
choice. If we were to leave the (singular) 
gauge for the (anti)instanton, in which the solutions of Eq.\,(\ref{HS}) are constructed, 
by a time-dependent gauge rotation $\bar{\Omega}
(\tau)$ then a pure-gauge configuration 
$A^{p.g.}_\mu(\tau)=i\delta_{\mu4}\bar{\Omega}^\dagger\pd_{\tau}\bar{\Omega}$ 
would appear in a {\sl covariant} derivative on the left-hand side 
of Eq.\,(\ref{BPSphi}).}, $V\equiv\mbox{tr}\,\left(V^{(1/2)}\right)^\dagger\,V^{(1/2)}$. 
In Eq.\,(\ref{BPSphi}) the right-hand side will turn out to be determined only up to 
a global gauge rotation, see Fig.\,\ref{winding}.
%***********************
%************************ 
Already at this point it is important to remark
that the Yang-Mills scale parametrizes the potential 
$V$ and thus also the classical solution to Eq.\,(\ref{BPSphi}). 
In the absence of trivial-topology 
fluctuations it is, however, invisible, 
see Eq.\,(\ref{thetamacro}). Only after the 
macroscopic equation of motion for the trivial-topology sector is 
solved for a pure-gauge configuration in the background $\phi$ 
does the existence of a Yang-Mills 
scale become manifest by a nonvanishing ground-state 
pressure and a nonvanishing ground-state 
energy density \cite{Hofmann2004}. Hence the trace anomaly $\tilde{\theta}_{\mu\mu}\not=0$
for the total energy-momentum tensor 
$\tilde{\theta}_{\mu\nu}\equiv \bar{\theta}^{g.s.}_{\mu\nu}+
\theta^{\tiny\mbox{fluc}}_{\mu\nu}$ in the effective theory
which includes the effects of 
trivial-topology fluctuations: 
Since $\bar{\theta}^{g.s.}_{\mu\nu}=4\pi\,T\Lambda^3\,\delta_{\mu\nu}$ and 
$\theta^{\tiny\mbox{fluc}}_{\mu\nu}\propto T^4$ for $T\gg \Lambda$ the trace 
anomaly dies off as $\frac{\Lambda^3}{T^3}$. 

\noindent Without imposing constraints other than nonlocality 
\footnote{A local definition of $\phi$'s phase 
always yields zero due to the (anti)selfduality of the (anti)caloron configuration.} 
the $\tau$ dependence of $\phi$'s phase (the ratio of the two averages $\phi$ and $|\phi|$ 
over admissible moduli defomrmations of $A^(C,A)$) would naively be characterized as
%*********
\eab
\label{defphi}
\frac{\phi^a}{|\phi|}(\tau)&\sim &\mbox{tr}\Bigg[\nonumber\\ 
&&\beta^0 1!\int d^3x\,\int d\rho\, \nonumber\\ 
&&\frac{\lambda^a}{2} F_{\mu\nu}[A_\alpha(\rho,\beta)]\left((\tau,0)\right)\,
\left\{(\tau,0),(\tau,\vec{x})\right\}[A_\alpha(\rho,\beta)]\times
\nonumber\\ 
&&F_{\mu\nu}[A_\alpha(\rho,\beta)]\left((\tau,\vec{x})\right)\,
\left\{(\tau,\vec{x}),(\tau,0)\right\}[A_\alpha(\rho,\beta)]+
\nonumber\\ 
&&\beta^{-1} 2!\int d^3x\int d^3y\,\int d\rho\, 
\nonumber\\ 
&&\frac{\lambda^a}{2} F_{\mu\lambda}[A_\alpha(\rho,\beta)]\left((\tau,0)\right)\,
\left\{(\tau,0),(\tau,\vec{x})\right\}[A_\alpha(\rho,\beta)]\times
\nonumber\\ 
&&\,F_{\lambda\nu}[A_\alpha(\rho,\beta)]\left((\tau,\vec{x})\right)\,
\left\{(\tau,\vec{x}),(\tau,\vec{y})\right\}[A_\alpha(\rho,\beta)]\times
\nonumber\\ 
&&F_{\nu\mu}[A_\alpha(\rho,\beta)]\left((\tau,\vec{y})\right) \left\{(\tau\vec{y}),(\tau,0)\right\}+
\nonumber\\ 
&&\beta^{-2} 3!\int d^3x\,\int d^3y\,\int d^3u\,\int d\rho\, 
\nonumber\\ 
&&\frac{\lambda^a}{2} F_{\mu\lambda}[A_\alpha(\rho,\beta)]\left((\tau,0)\right)\,
\left\{(\tau,0),(\tau,\vec{x})\right\}[A_\alpha(\rho,\beta)]\times
\nonumber\\ 
&&\,F_{\lambda\nu}[A_\alpha(\rho,\beta)]\left((\tau,\vec{x})\right)\,
\left\{(\tau,\vec{x}),(\tau,\vec{y})\right\}[A_\alpha(\rho,\beta)]\times
\nonumber\\ 
&&F_{\nu\kappa}[A_\alpha(\rho,\beta)]\left((\tau,\vec{y})\right) 
\left\{(\tau\vec{y}),(\tau,\vec{u})\right\}
F_{\kappa\mu}[A_\alpha(\rho,\beta)]\left((\tau,\vec{u})\right)\times 
\nonumber\\ 
&&\left\{(\tau,\vec{u}),(\tau,0)\right\}+\cdots\Bigg]\,.
\eae
%*********
The dots in (\ref{defphi}) stand for the contributions of 
higher $n$-point functions and for reducible, 
that is factorizable, contributions with respect to the spatial integrations.

\noindent In (\ref{defphi}) the following definitions apply: 
%*********
\eab
\label{defdefphi}
|\phi|&\equiv&\frac{1}{2}\,\mbox{tr}\,\phi^2\,
,\nonumber\\ 
\left\{(\tau,0),(\tau,\vec{x})\right\}[A_\alpha]&\equiv& {\cal P}\,
\exp\left[i\int_{(\tau,0)}^{(\tau,\vec{x})}dy_\beta\,A_\beta(y,\rho)\right]
\,,\nonumber\\ 
\left\{(\tau,\vec{x}),(\tau,0)\right\}[A_\alpha]&\equiv& {\cal P}\,
\exp\left[-i\int_{(\tau,0)}^{(\tau,\vec{x})}dy_\beta\,A_\beta(y,\rho)\right]\,.
\eae
%*********
The integral in the Wilson lines in Eqs.\,(\ref{defdefphi}) 
is along a straight line\footnote{Curved integration contours
introduce scales which have no physical counterpart on the classical level.
Furthermore, shifting the spatial part of the argument 
$(\tau,0)\to(\tau,\vec{z})$ in (\ref{defphi}) 
introduces a parameter $|\vec{z}|$ of dimension inverse mass.
There is no physical reason for a finite value of $|\vec{z}|$ to exist on the classical level.
Thus we conclude that $\vec{z}=0$.}
connecting the points $(\tau,0)$ and $(\tau,\vec{x})$, and ${\cal P}$ 
denotes the path-ordering symbol. 

\noindent Under a microscopic gauge transformation $\Omega(y)$ the following 
relations hold:
%*********
\eab
\label{micrOm}
\left\{(\tau,0),(\tau,\vec{x})\right\}[A_\alpha]&\rightarrow & \Omega^\dagger ((\tau,0))\,
\left\{(\tau,0),(\tau,\vec{x})\right\}[A_\alpha]\,
\Omega((\tau,\vec{x}))\,,\nonumber\\ 
\left\{(\tau,\vec{x}),(\tau,0)\right\}[A_\alpha]&\rightarrow & \Omega^\dagger 
((\tau,\vec{x}))\,\left\{(\tau,\vec{x}),(\tau,0)\right\}[A_\alpha]\,
\Omega((\tau,0))\,,\nonumber\\ 
F_{\mu\nu}[A_\alpha]\left((\tau,\vec{x})\right)&\rightarrow &
\Omega^\dagger((\tau,\vec{x}))\,F_{\mu\nu}[A_\alpha]((\tau,\vec{x}))\,\Omega((\tau,\vec{x}))\,,\nonumber\\ 
F_{\mu\nu}[A_\alpha]\left((\tau,0)\right)&\rightarrow &
\Omega^\dagger((\tau,0))\,F_{\mu\nu}[A_\alpha]((\tau,0))\,\Omega((\tau,0))\,.
\eae
%*********
As a consequence of Eq.\,(\ref{micrOm}) the right-hand side of (\ref{defphi}) transforms as 
%*********
\eqb
\label{phitrans}
\frac{\phi^a}{|\phi|}(\tau)\rightarrow R_{ab}(\tau)\,\frac{\phi^b}{|\phi|}(\tau)
\eqe
%********
where the SO(3) matrix $R_{ab}(\tau)$ is defined as
%*******
\eqb
\label{Rdef}
R^{ab}(\tau)\lambda^b=\Omega((\tau,0))\,\lambda^a\,\Omega^\dagger((\tau,0))\,.
\eqe
%******
Thus we have defined an adjointly transforming 
scalar in (\ref{defphi}). Moreover, only the time-dependent part of a microscopic gauge transformation 
survives after spatial averaging (macroscopic level).

In (\ref{defphi}) the $\sim$ sign indicates that both left- and right-hand sides satisfy the same 
homogeneous evolution equation in $\tau$  
%**********
\eqb
\label{deffequationhomo}
{\cal D}\left[\frac{\phi}{|\phi|}\right]=0\,.
\eqe
%*********
Here ${\cal D}$ is 
a differential operator such that Eq.\,(\ref{deffequationhomo}) represents a 
homogeneous differential equation. As it will turn out, Eq.\,(\ref{deffequationhomo}) is a {\sl linear} 
second-order equation which, up to global gauge rotations, 
determines the first-order or BPS equation whose solution $\phi$'s phase is. 
Each term in the series in (\ref{defphi}) is understood as a sum 
over the two solutions in Eq.\,(\ref{HS}), that is, $A_\alpha=
A_\alpha^C$ or $A_\alpha=A_\alpha^A$. 
As we shall show in Sec.\,\ref{comphase}, the dimensionless quantity defined on the right-hand side of (\ref{defphi}) 
is ambiguous\footnote{A shift $\tau \to \tau + \tau_z$\, ($0\le \tau_z \le \beta$),
is always possible and not fixed by a physical boundary condition on the classical level.
As we shall see below, the same holds true for the normalization of the right-hand side of (\ref{defphi}).
Therefore, for each color direction these two ambiguities parametrize the solution space of the second-order linear operator
${\cal D}$.}, the operator ${\cal D}$, however, is not.

The quantities appearing in the numerator and denominator of the left-hand side of
(\ref{defphi}) are understood as functional and spatial averages over the appropriate multilocal operators, 
being built of the field strength and the gauge field. The functional average is 
restricted to the moduli spaces of $A_\alpha=A_\alpha^C$ and $A_\alpha=A_\alpha^A$ 
excluding global color rotations and time translations. 

Let us explain this in more detail.
For the gauge variant 
density in (\ref{defphi}) an average over global 
color rotations and time shifts $\tau \to \tau + \tau_z$ 
($0\le \tau_z \le \beta$) would yield zero and 
thus is forbidden\footnote{The 'naked' gauge charge in (\ref{defphi}) is needed for a coupling 
of the trivial topology sector to the ground-state after spatial coarse-graining 
generating (i) quasiparticle masses and 
(ii) finite values of the ground-state energy density and the ground-state 
pressure \cite{Hofmann2004}.}. The nonflatness of the measure with respect to the separate $\rho$ 
integration in the numerator and the denominator average in (\ref{defphi}) transforms into 
flatness by taking the ratio. 
Since the integration 
weight $\exp(-S)$ is independent of temperature on the moduli space of a caloron 
the right-hand side of (\ref{defphi}) must not exhibit an explicit temperature dependence.
This forbids the contribution of $n$-point functions with $n>2$,
and we are left with an investigation of the first term in (\ref{defphi}). 
In the absence of a fixed mass scale 
on the classical level an average over 
spatial translations would have a dimensionful measure $d^3z$ making the definition of a dimensionless
quantity $\sim\frac{\phi}{|\phi|}$ impossible. We conclude that the 
average over spatial translations is already performed in (\ref{defphi}). 
Since one of the two available length 
scales $\rho$ and $\beta$ parametrizing the caloron or the anticaloron 
is integrated over in (\ref{defphi}) the only scale 
responsible for a nontrivial $\tau$ 
dependence of $\frac{\phi^a}{|\phi|}$ is $\beta$. 

What about the contribution of calorons with a topological charge modulus larger than unity? 
Let us consider the charge-two case.
Here we have three moduli of dimension length
which should enter the average defining the differential operator $\cal D$: 
two scale parameters and a core separation.
The reader may easily convince himself 
that by the absence of an explicit temperature dependence 
it is impossible to define the associated dimensionless quantity
in terms of spatial and moduli averages over $n$-point functions involving these configurations.
The situation is even worse for calorons of topological charge larger than two.
We conclude that only calorons of topological charge $\pm 1$ contribute to the 
definition of the operator $\cal D$ in Eq.\,(\ref{deffequationhomo}) 
by means of Eq.\,(\ref{defphi}).

\section{Computation of two-point correlation\label{comphase}}

Before we perform the actual calculation let us stress some simplifying 
properties of the solutions $A^{(C,A)}_\mu$ in Eq.\,(\ref{HS}). 

The path-ordering prescription for the 
Wilson lines $\left\{(\tau,0),(\tau,\vec{x})\right\}$ and 
$\left\{(\tau,0),(\tau,\vec{x})\right\}^\dagger$ in 
Eq.\,(\ref{defdefphi}) can actually be omitted.
To see this, we first consider 
the quantity $P^{(C,A)}(\tau,rs\vec{t})$ defined as
%***********
\eqb
\label{scprod}
P^{(C,A)}(\tau,rs\vec{t})\equiv A^{(C,A)}_i(\tau,sr\vec{t})\,t_i\,
\eqe
%***********
where $0\le r\le \infty$, $0\le s\le 1$, $(i=1,2,3)$. The vector $\vec{t}$ denotes the 
unit line-tangential along the straight line connecting 
the points $(\tau,0)$ and 
$(\tau,\vec x)\equiv (\tau,r\vec{t})$. We have 
%**********
\eqb
\label{WP}
\left\{(\tau,0),(\tau,r\vec{t}\right)\}^{(C,A)}=
{\cal P}\exp\left[i r\int_0^1ds\,P^{(C,A)}(\tau,sr\vec{t})\right]\,
\eqe
%********
where 
%**********
\eqb
\label{PHH}
P^{(C,A)}(\tau,sr\vec{t})=\mp \frac12\,\vec{t}\cdot\boldsymbol
{\lambda}\,\partial_4\ln\Pi(\tau,sr)\,.
\eqe
%********
Thus the path-ordering symbol can, indeed, be omitted in Eq.\,(\ref{WP}).
The field strength $F^C_{a\mu\nu}$ on the caloron solution 
in Eq.\,(\ref{HS}) is
%************
\eab \label{fieldstrength}
F^C_{a\mu\nu} &=& \bar\eta_{a\mu\nu} \frac{\left( \partial_{\kappa} \Pi \right) \left(\partial_{\kappa} \Pi \right)}{\Pi^2}
+ \bar\eta_{a\mu\kappa} \frac{\Pi \left( \partial_{\nu}\partial_{\kappa} \Pi \right) - 2 \left( \partial_{\kappa} \Pi
\right) \left(\partial_{\nu} \Pi \right)}{\Pi^2}  \nonumber \\ 
&& - \bar\eta_{a\nu\kappa} \frac{\Pi \left( \partial_{\mu}\partial_{\kappa} \Pi \right) - 2 \left( \partial_{\kappa} \Pi
\right) \left(\partial_{\mu} \Pi \right)}{\Pi^2}\,
\eae
%********
where $\Pi$ is defined in Eq.\,(\ref{Pi}). 
For the anticaloron one replaces $\bar\eta$ by $\eta$ in
Eq.\,(\ref{fieldstrength}). Using Eqs.\,(\ref{defphi}), (\ref{WP}), and (\ref{fieldstrength}), 
we obtain the following expression for the contribution 
$\left.\frac{\phi^a}{|\phi|}\right|_{\tiny{C}}$ arising from calorons:
%***********
\eab
\label{3DUP}
\left.\frac{\phi^a}{|\phi|}\right|_{\tiny{C}}&\sim&
i\int d\rho \int d^3x\,\frac{x^a}{r}
\left[\frac{\left(\partial_4\Pi(\tau,0)\right)^2}{\Pi^2(\tau,0)}-\frac23
\frac{\partial^2_4\Pi(\tau,0)}{\Pi(\tau,0)}\right]\times\nonumber\\  
&&\left\{4\cos(2g(\tau,r))\left[\frac{\pd_r\pd_4\Pi(\tau,r)}{\Pi(\tau,r)}-
2\frac{\left(\pd_r\Pi(\tau,r)\right)\left(\pd_4\Pi(\tau,r)\right)}
{\Pi^2(\tau,r)}\right]+\right.\nonumber\\ 
&&\left.\sin(2g(\tau,r))\left[4\,\frac{\left(\pd_4\Pi(\tau,r)\right)^2-
\left(\pd_r\Pi(\tau,r)\right)^2}{\Pi^2(\tau,r)}+
2\,\frac{\pd^2_r\Pi(\tau,r)-\pd^2_4\Pi(\tau,r)}{\Pi(\tau,r)}
\right]\right\}\, \nonumber\\ 
\eae
%********** 
where 
%********
\eqb
\label{grtau}
g(\tau,r)\equiv \int_0^1 ds\, \frac{r}{2} \, \pd_4 \ln \Pi(\tau,sr)
\eqe
%******** 
and 
%********
\eqb
\label{beiNull}
\frac{\left(\partial_4\Pi(\tau,0)\right)^2}{\Pi^2(\tau,0)}-\frac23
\frac{\partial^2_4\Pi(\tau,0)}{\Pi(\tau,0)}=
-\frac{16}{3}\pi^4\frac{\rho^2}{\beta^2} 
\frac{\pi^2 \rho^2 + \beta^2 \left(2 + \cos \left(\frac{2\pi\tau}{\beta}\right)\right)}
{\left[2\pi^2 \rho^2 + \beta^2 \left(1-\cos \left(\frac{2\pi\tau}{\beta}\right)\right)\right]^2}\,.
\eqe
%********
The dependences on $\rho$ and
$\beta$ are suppressed in the integrands of (\ref{3DUP}) 
and Eq.\,(\ref{grtau}). It is worth mentioning 
that the integrand in Eq.\,(\ref{grtau}) is proportional to $\delta(s)$ 
for $r\gg\beta$. A useful set of identities is
%*********
\eab
\label{inversion}
F_{\mu\nu}^C(\tau,\vec{x}) &=& F_{\mu\nu}^A(\tau,-\vec{x}) \nonumber\\
\{(\tau,0),(\tau,\vec{x})\}^C &=& \left(\{(\tau,\vec{x}),(\tau,0)\}^C\right)^{\dagger} = \nonumber\\
\{(\tau,0),(\tau,-\vec{x})\}^A &=& \left(\{(\tau,-\vec{x}),(\tau,0\}^A\right)^{\dagger} \,.
\eae
%*********
Eqs.\,(\ref{inversion}) state that the integrand for 
$\left.\frac{\phi^a}{|\phi|}\right|_{\tiny{A}}$ can be obtained 
by a parity transformation $\vec{x}\to-\vec{x}$ of the integrand for 
$\left.\frac{\phi^a}{|\phi|}\right|_{\tiny{C}}$. Since the latter changes 
its sign, see (\ref{3DUP}), one naively would conclude that 
%********
\eqb
\label{naivcanc}
\frac{\phi^a}{|\phi|}=
\left.\frac{\phi^a}{|\phi|}\right|_{\tiny{C}}+ 
\left.\frac{\phi^a}{|\phi|}\right|_{\tiny{A}}=0\,.
\eqe
%*********
This, however, would only be the case if no ambiguity in evaluating the 
integral in both cases existed. But such ambiguities do occur! 
First, the $\tau$ dependence of the anticaloron's contribution may be shifted 
as compared to that of the caloron.
Second, the color orientation of caloron and anticaloron contributions may be different.
Third, the normalization of the two contributions may be different.
To see this, we need to investigate the convergence properties of the radial 
integration in (\ref{3DUP}). It is easily checked that 
all terms give rise to a converging $r$ integration 
except for the following one:
%***********
\eqb
\label{nonconvr}
2\,\frac{x^a}{r}\,\sin(2g(\tau,r))\,\frac{\pd^2_r\Pi(\tau,r)}{\Pi(\tau,r)}\,.
\eqe
%***********
Namely, for $r>R\gg\beta$ (\ref{nonconvr}) goes over in
%***********
\eqb
\label{nonconvrlarger}
4\,t^a\frac{\pi\rho^2\sin(2g(\tau,r))}{\beta r^3}\,.
\eqe
%***********
Thus the $r$-integral of the term in (\ref{nonconvr}) is 
logarithmically divergent in the infrared\footnote{The integral converges for $r\to 0$.}:
%***********
\eqb
\label{nocI}
4\,t^a\,\frac{\pi\rho^2}{\beta}\int_R^\infty \frac{dr}{r}\,
\sin(2g(\tau,r))\,\,.
\eqe
%***********
Recall that $g(\tau,r)$ behaves like a constant 
in $r$ for $r>R$. The angular 
integration, on the other hand, would 
yield zero if the radial integration was regular. 
Thus a logarithmic divergence can be cancelled 
by the vanishing angular integral to yield some 
finite and real but undetermined normalization of the emerging $\tau$ dependence. 
To investigate this, both angular and radial 
integration need to regularized.     

\noindent We may regularize the $r$ integral in (\ref{nocI}) by prescribing
%************
\eqb
\label{Glargerreg}
\int_R^\infty \frac{dr}{r}
\to
\beta^\epsilon \int_R^\infty\frac{dr}{r^{1+\epsilon}}\,. 
\eqe
%************* 
with $\epsilon>0$. We have
%************
\eab
\label{regr}
\beta^\epsilon\int_R^\infty \frac{dr}{r^{1+\epsilon}}&=&
\beta^\epsilon\int_0^\infty \frac{dr}{(r+R)^{1+\epsilon}}\nonumber\\ 
&=&\frac{1}{\epsilon}-\log\left(\frac{R}{\beta}\right)+\frac{1}{2}\,
\epsilon\log^2\left(\frac{R}{\beta}\right)+\cdots\,.
\eae
%*************
Away from the pole at $\epsilon=0$ this is regular. For $\epsilon<0$ 
Eq.\,(\ref{regr}) can be regarded as a legitimate analytical 
continuation. An ambiguity inherent in Eq.\,(\ref{regr}) 
relates to how one circumvents the pole in the 
smeared expression
%************
\eab
\label{smearing}
&&\frac{1}{2 \eta}\int_{-\eta}^{\eta}d\epsilon \left(\frac{1}{\epsilon\pm 
i0}-\log\left(\frac{R}{\beta}\right)+\frac{1}{2}\,\epsilon\log^2
\left(\frac{R}{\beta}\right)+\cdots\right)\nonumber\\ 
&=&\mp\frac{\pi i}{2 \eta}-\log\left(\frac{R}{\beta}\right)+\cdots\,,\ \ \ (\eta>0\,,\eta\ll 1)\,.
\eae
%***********
Concerning the regularization of the angular 
integration we may introduce defect (or surplus) angles 
$2\eta^\prime$ in the azimuthal integration as
%************
\eqb
\label{defect}
\int_0^\pi d\omega\,\sin\omega\int_0^{2\pi}d\theta\to \int_{0}^
{\pi} d\omega\,\sin\omega\int_{\alpha_C\pm\eta^\prime}^{\alpha_C+2\pi\mp\eta^\prime}d\theta \,.
\eqe
%***********
In Eq.\,(\ref{defect}) $\alpha_C$ is a constant angle with $0\le\alpha_C\le 2\pi$ 
and $0<\eta^\prime\ll 1$.   
%***********************
\begin{figure}
\begin{center}
%\leavevmode
%%\epsfxsize=9.cm
%\leavevmode
%%\epsffile[80 25 534 344]{}
\vspace{5.3cm}
\includegraphics{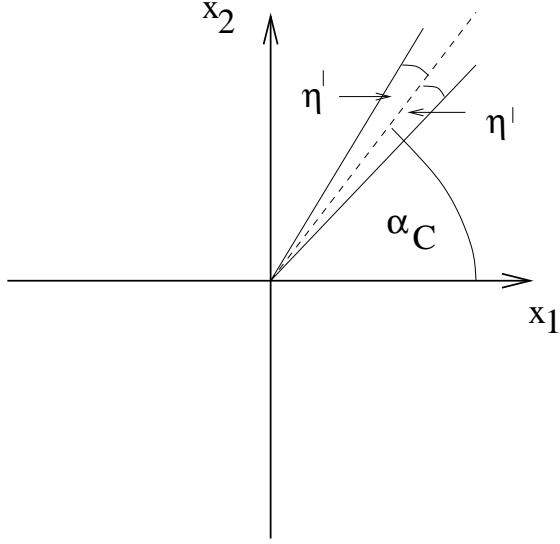}
\end{center}
\caption{The axis for the integration over $\theta$.\label{Fig6}}      
\end{figure}
%************************
Obviously, this regularization singles out the $x_1x_2$ 
plane. As we shall show below, the choice of regularization 
plane translates into a global gauge choice 
for the $\tau$ dependence of $\phi$'s phase and thus 
is physically irrelevant: The apparent breaking of rotational 
symmetry by the angular regularization translates into a gauge choice.  

The value of $\alpha_C$ is 
determined by a (physically irrelevant)
initial condition. 
We have
%*********
\eqb
\label{regangint}
\int_{0}^
{\pi} d\omega\,\sin\omega
\int_{\alpha_C\pm\eta^\prime}^{\alpha_C+2\pi\mp\eta^\prime}d\theta\, t^a\sim \mp
\pi\eta^\prime\left(\delta_{a1}\cos\alpha_C+\delta_{a2}\sin\alpha_C\right)\,.
\eqe
%**********
To see what is going on we may fix, for the time being, 
the ratio $\frac{\eta^\prime}{\eta}$ for the normalization of the caloron contribution
to a finite and positive but otherwise arbitrary constant $\Xi$ when sending ${\eta}$ and $\eta^\prime$ 
to zero in the end of the calculation:
%********
\eqb\label{Xi}
\lim_{\eta,\,\eta^\prime\to0} \,\,\frac{\eta^\prime}{\eta} = \Xi \,.
\eqe
%**********
Combining Eqs.\,(\ref{nocI}),(\ref{smearing}),(\ref{regangint}), (\ref{beiNull}), 
expression (\ref{3DUP}) reads: 
%**********
\eab
\label{combined}
\left.\frac{\phi^a}{|\phi|}\right|_{\tiny{C}}&\sim&
\pm\frac{32}{3}\,\frac{\pi^7}{\beta^3}\,\Xi\,\left(\delta_{a1}\cos\alpha_C+\delta_{a2}\sin\alpha_C\right)\,\int d\rho\,\left[\lim_{r\to\infty}\sin(2g(\tau,r))\right]\times\nonumber\\ 
&&\rho^4\frac{\pi^2 \rho^2 + \beta^2 \left(2 + \cos \left(\frac{2\pi\tau}{\beta}\right)\right)}
{\left[2\pi^2 \rho^2 + \beta^2 \left(1-\cos \left(\frac{2\pi\tau}{\beta}\right)\right)\right]^2}\nonumber\\ 
&\equiv&\pm\Xi\,\left(\delta_{a1}\cos\alpha_C+\delta_{a2}\sin\alpha_C\right) {\cal A}
\left(\frac{2\pi\tau}{\beta}\right)\,
\eae
%********** 
where ${\cal A}$ is a dimensionless function of its dimensionless argument. 
The sign ambiguity in (\ref{combined}) arises from the 
ambiguity associated with the way how one circumvents the pole 
in Eq.\,(\ref{smearing}) and whether one introduces a surplus or a defect angle
in (\ref{defect}). 
Furthermore, there is an ambiguity associated with a constant shift $\tau\to\tau+\tau_C$
($0\le\tau_C\le\beta$) in Eq.\,(\ref{combined}).

For the anticaloron contribution we may, for the time being, fix the ratio $\frac{\eta^\prime}{\eta}$
to another finite and positive constant $\Xi'$.
In analogy to the caloron case, there is the ambiguity related to a shift $\tau\to\tau+\tau_A$ ($0\le\tau_A\le\beta$)
in the anticaloron contribution.
Moreover, we may without restriction of generality (global gauge choice)
use an axis for the angular regularization which also lies in the
$x_1x_2$ plane, but with a different angle $\alpha_A$.
Then we have
%*********
\eab
\label{totalc}
\frac{\phi^a}{|\phi|}
&=&\left.\frac{\phi^a}{|\phi|}\right|_{\tiny{C}}+
\left.\frac{\phi^a}{|\phi|}\right|_{\tiny{A}}\nonumber\\ 
&=&
\pm\Xi\,\left(\delta_{a1}\cos\alpha_C+\delta_{a2}\sin\alpha_C\right) {\cal A}
\left(\frac{2\pi(\tau+\tau_C)}{\beta}\right)
\nonumber\\ 
&&\pm\Xi'\,\left(\delta_{a1}\cos\alpha_A+\delta_{a2}\sin\alpha_A\right) {\cal A}
\left(\frac{2\pi(\tau+\tau_A)}{\beta}\right)
\nonumber\\ 
&\neq&0 
\eae
%*********
where the choices of signs in either contribution are independent.
Eq.\,(\ref{totalc}) is the basis for fixing the operator $\cal D$ in Eq.\,(\ref{deffequationhomo}).

\begin{figure}
\begin{center}
%\leavevmode
%%\epsfxsize=9.cm
%\leavevmode
%%\epsffile[80 25 534 344]{}
\vspace{5.3cm}
\includegraphics{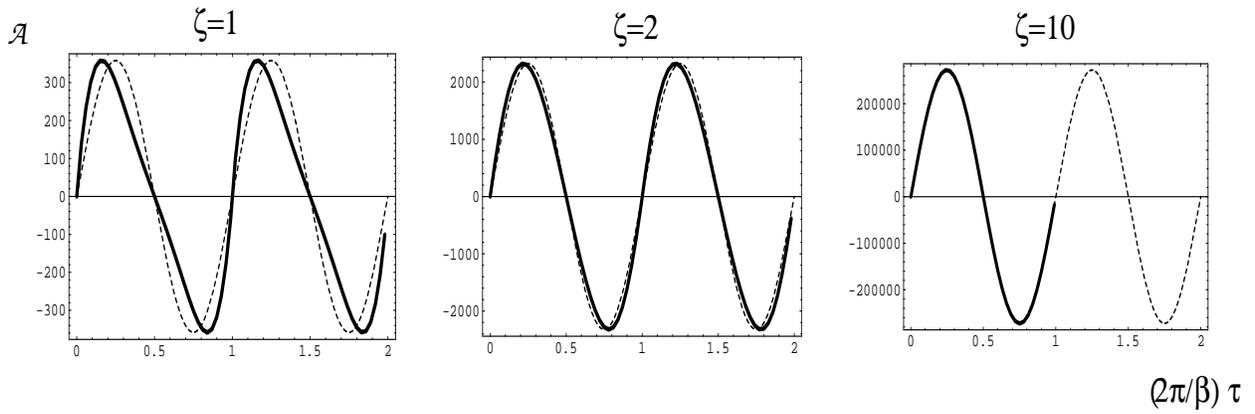}
\end{center}
\caption{${\cal A}$ as a function of $\frac{2\pi}{\beta}\tau$ for $\zeta=1,2,10$. For each case 
the dashed line is a plot of $\mbox{max}\,{\cal A}\times\sin\left(\frac{2\pi}{\beta}\tau\right)$. 
We have fitted the asymptotic dependence on $\zeta$ of the amplitude of ${\cal A}$ as 
${\cal A}\left(\frac{2\pi}{\beta}\tau=\frac{\pi}{2},\zeta\right)=272\,\zeta^3,\,(\zeta>10)$. 
The fit is stable under variations of the
fitting interval. For the case $\zeta=10$ the difference between 
the two curves can not be resolved anymore.\label{Fig7}}      
\end{figure}

\noindent To evaluate the function ${\cal A}\left(\frac{2\pi\tau}{\beta}\right)$ 
in Eq.\,(\ref{combined}) numerically, we introduce the same cutoff for the $\rho$ integration 
in the caloron and anticaloron case 
as follows:
%********
\eqb
\label{cutoffrho}
\int d\rho\to \int_0^{\zeta \beta} d\rho\,,\ \ \ \ \ \ (\zeta>0)\,.
\eqe
%*********
This introduces an additional dependence of ${\cal A}$ on $\zeta$. 
In Fig.\,\ref{Fig7} the $\tau$ dependence of ${\cal A}$ for 
various values of $\zeta$ is depicted. 
%***********************
%************************
It can be seen that 
%**********
\eqb
\label{zetas}
{\cal A}\left(\frac{2\pi}{\beta}\tau,\zeta\to\infty\right)\to 
272\,\zeta^3\,\sin\left(\frac{2\pi}{\beta}\tau\right)\,.
\eqe
%**********
Therefore we have 
%*********
\eab
\label{finalres}
\frac{\phi^a}{|\phi|} &\sim& 
272\,\zeta^3\,
\Bigg(
\pm \Xi 
\left(\delta_{a1}\cos\alpha_C+\delta_{a2}\sin\alpha_C\right)\,
\sin\left(\frac{2\pi}{\beta}(\tau+\tau_C)\right)
\nonumber \\ &&
\pm \Xi' 
\left(\delta_{a1}\cos\alpha_A+\delta_{a2}\sin\alpha_A\right)\,
\sin\left(\frac{2\pi}{\beta}(\tau+\tau_A)\right)
\Bigg)
\nonumber\\
&\equiv& \hat \phi^a
\,.
\eae
%********** 
The numbers $\zeta^3\,\Xi$, $\zeta^3\,\Xi'$, $\frac{\tau_C}{\beta}$ and $\frac{\tau_A}{\beta}$ 
in (\ref{finalres}) are undetermined. 
For each color orientation (corresponding to a given angular regularization)
there are two independent parameters,
a normalization and a phase-shift.
The principal impossibility to fix the normalizations
reflects the fact that on the classical 
level the theory is invariant under spatial dilatations. To give a meaning to these number, 
a mass scale needs to be generated dynamically. This, however, 
can only happen due to dimensional transmutation, which is known to be an effect 
induced by trivial-topology fluctuations \cite{NP2004}.
The result in (\ref{finalres}) is highly nontrivial since it is obtained only after
an integration over the entire admissible part of the moduli spaces of (anti)calorons is performed.

Let us now discuss the physical content of (\ref{finalres}). For fixed values of
the parameters $\zeta^3\,\Xi$, $\zeta^3\,\Xi'$, $\frac{\tau_C}{\beta}$ and $\frac{\tau_A}{\beta}$ 
the right-hand side of Eq\,(\ref{finalres}) resembles a fixed 
elliptic polarization in the $x_1x_2$ plane of adjoint color space. 
For a given polarization plane the two independent numbers (normalization and phase-shift) 
of each oscillation axis parametrize the solution space (in total four undetermined parameters) 
of a second-order linear differential equation 
\eqb\label{2order}
{\cal D} \hat\phi=0\,.
\eqe
From (\ref{finalres}) we observe that the operator $\cal D$ is
\eqb
{\cal D} = \partial_\tau^2 + \left( \frac{2\pi}{\beta} \right)^2\,.
\eqe
Since for a given polarization plane there is a one-to-one map from the solution space of Eq.\,(\ref{2order}) 
to the parameter space associated with the ambiguities 
in the definition (\ref{defphi}) 
we conclude that the operator ${\cal D}$ is {\sl uniquely} determined by (\ref{defphi}).   

What we need to assure the 
validity of Eq.\,(\ref{thetamacro}) is a BPS saturation\footnote{
The modulus $|\phi|$ does not depend on $\tau$ in thermal equilibrium and thus can be cancelled out.} 
of the solution to Eq.\,(\ref{2order}).
Thus we need to find first-order equations whose solutions solve the second-order equation (\ref{2order}).
The relevant two first-order equations are 
%***********
\eqb\label{1orderbps}
\pd_\tau \hat{\phi}=\pm\frac{2\pi i}{\beta}\lambda_3\,\hat{\phi}\,
\eqe
%**********
where we have defined $\phi=|\phi|\hat{\phi}(\tau)$. 
Obviously, the right-hand sides of Eqs.\,(\ref{1orderbps}) are subject to a 
global gauge ambiguity associated with the choice of plane for angular regularization,
any normalized generator other than $\lambda_3$ could have appeared, see Fig.\,\ref{winding}. 
Now the solution to either of the two equations (\ref{1orderbps})
also solves Eq.\,(\ref{2order}),
\eab
\partial_\tau^2 \hat\phi &=&
\pm \frac{2\pi i}{\beta}\lambda_3\, \partial_\tau \hat \phi
\nonumber\\
&=& \frac{2\pi i}{\beta}\lambda_3\,\frac{2\pi i}{\beta}\lambda_3\,\hat\phi
\nonumber\\
&=& - \left( \frac{2\pi}{\beta} \right)^2\hat\phi
\,.
\eae
Traceless, hermitian solutions to Eqs.\,(\ref{1orderbps}) are given as
%***********
\eqb\label{solutiona}
\hat{\phi} = C \, \lambda_1 \, \exp\left(\mp \frac{2\pi i}{\beta} \lambda_3 (\tau - \tau_0) \right)
\eqe
%************
where $C$ and $\tau_0$ denote real integration constants which both are undetermined. Notice 
that the requirement of BPS saturation has reduced the number of 
undetermined parameters from four to two: an elliptic polarization in the $x_1x_2$ plane is cast 
into a circular polarization. Thus the field $\phi$ winds along an $S^1$
on the group manifold $S^3$ of SU(2). Both winding senses appear but can 
not be distinguished physically \cite{Hofmann2004}.

\section{How to obtain $\phi$'s modulus\label{macro}}

Here we show how the information about $\phi$'s phase in Eq.\,(\ref{solutiona}) can be used 
to infer its modulus. 
Let us assume that a scale $\Lambda$ is externally given which 
characterizes this modulus at a given temperature $T$. Together, $\Lambda$ and $T$ 
determine what the minimal physical volume $|\phi|^{-3}$ is for which the spatial 
average over the caloron-anticaloron system saturates the infinite-volume average 
appearing in (\ref{defphi}).  

We have 
%************
\eqb
\label{tafel}
\phi = \phi\left(\beta, \Lambda,\frac{\tau}{\beta}\right)\,.
\eqe
%********
In order to reproduce the phase in Eq.\,(\ref{solutiona}) a {\sl linear} dependence on $\phi$
must appear on the right-hand side of the BPS equation (\ref{BPSphi}). Furthermore, 
this right-hand side ought not depend on $\beta$ explicitly and must be 
analytic in $\phi$\footnote{The former requirement derives from the fact that $\phi$ and its potential $V$ are 
obtained by functionally integrating over a noninteracting caloron-anticaloron system. 
The associated part of the partition function 
does not exhibit an explicit $\beta$ dependence since 
the action $S$ is $\beta$ independent on the caloron and anticaloron moduli spaces. 
Thus a $\beta$ dependence of $V$ or $V^{(1/2)}$ can only be generated 
via the periodicity of $\phi$ itself. The latter requirement 
derives from the demand that the thermodynamics at temperature $T + \delta T$ to any given accuracy 
must be derivable from the thermodynamics at temperature $T$ for $\delta T$ sufficiently 
small provided no phase transition occurs at $T$. This is accomplished by a Taylor expansion of 
the right-hand side of the
BPS equation (finite radius of convergence) which, in turn, is the starting point for a 
perturbative treatment with expansion parameter $\frac{\delta T}{T}$.}. 
The two following possibilities exist:
%*******
\eqb\label{bps13}
\pd_\tau\phi=\pm i\,\Lambda\,\lambda_3\,\phi 
\eqe
%**********
or 
%*******
\eqb\label{bps14}
\pd_\tau\phi=\pm i\,\Lambda^3\,\lambda_3\,\phi^{-1} 
\eqe
%**********
where $\phi^{-1}\equiv \frac{\phi}{|\phi|^2}$. 
Recall that
%***********
\eqb
\phi^{-1} = \phi_0^{-1} \sum_{n=0}^{\infty} (-1)^n \phi_0^{-n} \left(\phi-\phi_0\right)^n
\eqe
%*********
has a finite radius of convergence. 
According to Eq.\,(\ref{solutiona}) we may write
%*********
\eqb
\label{ansatzBPS}
\phi = |\phi|(\beta,\Lambda)\, \times\,\lambda_1\, \exp\left( \mp \frac{2\pi i}{\beta} \lambda_3 
(\tau - \tau_0)\right) \,.
\eqe
%***********
Substituting Eq.\,(\ref{ansatzBPS}) into 
Eq.\,(\ref{bps13}) yields
%*******
\eqb
\label{contraBPS}
\Lambda=\frac{2\pi}{\beta}
\eqe
%********
which is unacceptable since $\Lambda$ is a constant scale. For the possibility in
Eq.\,(\ref{bps14}) we obtain
%*******
\eqb
\label{nocontraBPS}
|\phi|(\beta,\Lambda)=\sqrt{\frac{\beta\Lambda^3}{2\pi}}=\sqrt{\frac{\Lambda^3}{2\pi\,T}}\,
\eqe
%********
when substituting Eq.\,(\ref{ansatzBPS}) into Eq.\,(\ref{bps14}). 
This is acceptable and indicates that at $T\gg \Lambda$ $\phi$'s modulus is small.
The right-hand side of Eq.\,(\ref{bps14}) defines the 'square-root' $V^{(1/2)}$ of a potential
$V(|\phi|)\equiv\mbox{tr}\,\left(V^{(1/2)}\right)^\dagger\,V^{(1/2)}=\Lambda^6 \, \mbox{tr} \, \phi^{-2}$,
and the equation of motion (\ref{bps14})
can be derived from the following action:
\eqb    \label{actionphi}
S_{\phi} = \mbox{tr} \, \int_0^\beta d\tau \int d^3x 
\left( \partial_\tau \phi \partial_\tau \phi + \Lambda^6 \phi^{-2}  \right)  \,.
\eqe 
Notice that a shift $V\to V+\mbox{const}$ is forbidden in Eq.\,(\ref{actionphi}) 
since the relevant equation of motion is the first-order equation (\ref{bps14}).

After the spatial average is performed the action $S_{\phi}$ is extended by including 
topologically trivial configurations $a_\mu$ in a minimal fashion: 
$\partial_\tau \phi \to \partial_\mu \phi + ie[\phi,a_\mu] \equiv D_\mu \phi$ and an added 
kinetic term. Here $e$ denotes the {\sl effective} gauge coupling.
Thus the effective Yang-Mills action $S$ is written as
\eqb \label{actiontotal}
S = \mbox{tr}\, \int_0^\beta d\tau \int d^3x \left( \frac12 G_{\mu\nu}G_{\mu\nu} +
D_\mu \phi D_\mu \phi + \Lambda^6 \phi^{-2} \right)
\,,
\eqe
where
$G_{\mu\nu}=G^a_{\mu\nu} \frac{\lambda^a}{2}$ and
$G^a_{\mu\nu}=\pd_\mu a^a_\nu-\pd_\nu a^a_\mu-e\epsilon^{abc}a^b_\mu a^c_\nu$.

In Eqs.\,(\ref{bps13}) and (\ref{bps14}) the existence of the mass scale $\Lambda$ (the Yang-Mills scale) 
was assumed. One attributes the generation of a mass scale to the topologically 
trivial sector which, however, was assumed to be switched off so far. How can a 
contradiction be avoided? The answer to this question is that the scale $\Lambda$ remains hidden as long as 
topologically trivial fluctuations are switched off, see Eq.\,(\ref{thetamacro}). 
Only after switching on interactions between trivial-holonomy calorons within the 
ground state can $\Lambda$ be seen \cite{Hofmann2004}. Let us repeat the 
derivation of this result: In \cite{Hofmann2004} we have shown 
that the mass-squared of $\phi$-field fluctuations, $\pd^2_{|\phi|}\,V(|\phi|)$, 
is much larger than the square of the compositeness 
scale $|\phi|$. Moreover $\pd^2_{|\phi|}\,V(|\phi|)$ is
much larger than $T^2$ for all temperatures $T\ge T_{c,E}$ where $T_{c,E}$ 
denotes the critical temperature for the
electric-magnetic transition. Thus $\phi$ is quantum mechanically 
and statistically inert: It provides a (nonbackreacting and 
undeformable) source for the following equation of motion 
%*********
\eqb
\label{gfeom}
{\cal D}_\mu G_{\mu\nu}=2ie[\phi,{\cal D}_\nu \phi] \,
\eqe
%**********
which follows from the action in Eq.\,(\ref{actiontotal}).
A pure-gauge solution to Eq.\,(\ref{gfeom}), describing the ground state together with $\phi$, is
%**********
\eqb
\label{curvfree}
a^{bg}_\mu=\frac{\pi}{e}\,T \delta_{\mu 4}\,\lambda_3\,.
\eqe
%*********** 
As a consequence of Eq.\,(\ref{curvfree}) we have ${\cal D}_{\mu} \phi \equiv 0$, 
and thus a ground-state pressure $P^{g.s}=-4\pi\,\Lambda^3\,T$ and a ground-state
energy-density $\rho^{g.s}=4\pi\,\Lambda^3\,T$ are generated in the electric 
phase: The so-far hidden scale $\Lambda$ becomes visible by averaged-over caloron-anticaloron 
interactions encoded in the pure-gauge configuration $a^{bg}_\mu$.

\section{Summary and Outlook\label{SO}}

Let us summarize our results. We have derived 
the phase and the modulus of a statistically and quantum mechanically inert adjoint and spatially homogeneous 
scalar field $\phi$ for an SU(2) Yang-Mills theory being in its electric phase. 
This field and a pure-gauge configuration together suggest the concept 
of a thermal ground state since they generate temperature 
dependent pressure and energy density with an equation of state corresponding to a 
cosmological constant. The existence of 
$\phi$ originates from the spatial correlations inherent in BPS saturated, trivial-holonomy 
solutions to the classical Yang-Mills equations at finite temperature: 
the Harrington-Shepard solutions of topological charge modulus one. To derive $\phi$'s phase 
these field configurations are, in a first step, treated as noninteracting when performing the 
functional average over the admissible parts of their moduli spaces. 
We have shown why adjoint scalar fields arising from 
configurations of higher topological charge do not exist. 

The BPS saturated and classical field $\phi$ possesses nontrivial 
$S^1$ winding on the group manifold $S^3$. The associated 
trajectory on $S^3$ becomes circular and thus a pure phase only after the 
integration over the entire admissible parts of the moduli spaces is carried out. 
Together with a pure-gauge configuration the adjoint scalar field
$\phi$ generates a linear temperature dependence of 
the ground-state pressure and the 
ground-state energy-density where the pure-gauge configuration 
solves the Yang-Mills equations in the 
background $\phi$ and, after the spatial average, describes the interactions
between trivial-holonomy calorons. 
The pure-gauge configuration also makes explicit that the electric phase is deconfining \cite{Hofmann2004}. 
Since trivial-topology 
fluctuations may acquire quasiparticle masses on tree-level by the 
adjoint Higgs mechanism \cite{Hofmann2004} 
the presence of $\phi$ resolves the infrared problem inherent 
in a perturbative loop expansion of 
thermodynamical quantities \cite{HerbstHofmannRohrer2004}. 
Since there are kinematical constraints for the maximal hardness 
of topologically trivial quantum fluctuations no renormalization procedure 
for the treatment of ultraviolet divergences is needed in the loop expansion of thermodynamical 
quantities \cite{HerbstHofmannRohrer2004} performed in the effective theory. 
These kinematical constraints arise from 
$\phi$'s compositeness emerging at distances $\sim |\phi|^{-1}$. 
The usual assertion that the effects of the topologically nontrivial sector are extremely suppressed 
at high temperature - they turn out to be {\sl power suppressed} in $T$ - is shown 
to be correct by taking this 
sector into account. The theory, indeed, has a Stefan-Boltzmann 
limit which is very quickly approached. It turns out to be incorrect, however, 
to neglect the topologically nontrivial sector from the start: assuming 
$T\gg\Lambda$ to justify the omission of the topologically nontrivial sector 
{\sl before} performing a (perturbative) loop expansion of thermodynamical quantities does 
not capture the thermodynamics of an SU(2) 
Yang-Mills theory and leads to the known problems 
in the infrared sector \cite{Linde1980}.

\section*{Acknowledgements}
We would like to thank Nucu Stamatescu for his 
continuing interest in our work, and Janos 
Polonyi and Dirk Rischke for stressing the necessity of this work. Useful conversations with 
H. Gies and J. Pawlowski are gratefully acknowledged.

\baselineskip25pt
\end{document}